\documentclass[%
 reprint,
 amsmath,amssymb,
 aps,
]{revtex4-1}

\usepackage{physics}
\usepackage{graphicx}
\usepackage{dcolumn}
\usepackage{bm}
\usepackage{xcolor}
\usepackage{amsmath}

\begin{document}


\title{On the preservation of coherence in the electronic wavepacket of a neutral and rigid polyatomic molecule}

\author{Andr\'{a}s Csehi}
 \affiliation{Department of Theoretical Physics, University of Debrecen, H-4002
Debrecen, PO Box 400, Hungary}

\author{P\'{e}ter Badank\'{o}}
 \affiliation{Department of Theoretical Physics, University of Debrecen, H-4002
Debrecen, PO Box 400, Hungary}

\author{G\'{a}bor J. Hal\'{a}sz}
 \affiliation{Department of Information Technology, University of Debrecen, H-4002
Debrecen, PO Box 400, Hungary}

\author{\'{A}gnes Vib\'{o}k}
 \email{vibok@phys.unideb.hu}
 \affiliation{Department of Theoretical Physics, University of Debrecen, H-4002
Debrecen, PO Box 400, Hungary}
 \affiliation{ELI-ALPS, ELI-HU Non-Profit Ltd, H-6720 Szeged, Dugonics t\'{e}r 13, Hungary}

\author{Benjamin Lasorne}
 \email{benjamin.lasorne@umontpellier.fr}
 \affiliation{ICGM, Univ Montpellier, CNRS, ENSCM, Montpellier, France}

\date{\today}

\begin{abstract}
We present various types of reduced models including five vibrational modes and three electronic states for the pyrazine molecule in order to investigate the lifetime of electronic coherence in a rigid and neutral system. 
Using an ultrafast optical pumping in the ground state ($1 ^{1}\mathrm{A_g}$), we prepare a coherent superposition of two bright excited states, $1 ^{1}\mathrm{B_{2u}}$ and $1 ^{1}\mathrm{B_{1u}}$, and reveal the effect of the nuclear motion on the preservation of the electronic coherence induced by the laser pulse.
More specifically, two aspects are considered: the anharmonicity of the potential energy surfaces and the dependence of the transition dipole moments (TDMs) with respect to the nuclear coordinates.
To this end, we define an “ideal model” by making three approximations: (i) only the five totally symmetric modes move, (ii) which correspond to uncoupled harmonic oscillators, and (iii) the TDMs from the ground electronic state to the two bright states are constant (Franck-Condon approximation). We then lift the second and third approximations by considering, first, the effect of anharmonicity, second, the effect of coordinate-dependence of the TDMs (first-order Herzberg-Teller contribution), third, both. 
Our detailed numerical study with quantum dynamics confirms long-term revivals of the electronic coherence even for the most realistic model.
\end{abstract}

\maketitle



\leftline {\bf I. INTRODUCTION}
\vspace{0.5cm}

Advances in laser technology have enabled the generation of virtually
arbitrary waveforms in the optical range \cite{krausz,Cundiff,vrakking1},
allowing one to shape oscillations of the electromagnetic field on
the sub-laser-cycle, attosecond time scale. Such pulses allow one
to take real-time snapshots of ultrafast transformations of matter
\cite{Vrakking2}. Pump-probe techniques using ultrashort sub-cycle
laser pulses have made it possible to control the motion of electrons
in atoms \cite{santra1,santra2,santra3}, molecules \cite{Bandrauk1,Bandrauk2,Kelkensberg,Fischer,Fernando1,Sansone,Steffi,Regina,Regina2,Regina3,Decleva,Bandrauk3,Palacios},
and condensed phase systems \cite{Reinhard1} such as dielectrics,
raising the fundamental question in this context about the relevance of the
time scale of electronic coherences for chemical dynamics evolving
on a much longer time scale: are electronic wavepackets created by
ultrashort pulses relevant for the making and breaking of chemical bonds in a molecule?

Experimental and theoretical works \cite{Tseng,Tom1,Tom2,Tom3,Sun,Banares}
on molecular fragmentation triggered by ionization suggest a positive
answer to this question. Sudden ionization of a molecule creates a coherent
superposition of electronic states, leading to ultrafast charge migration
\cite{lenz1,kuleff2,lenz2,calegari1,Kraus,kuleff3}. During this ultrashort
process one can gain insight into the electronic motion following
the ionization by describing the time evolution of the charge density
of the molecule. It is suggested that the coupling of this electronic
motion to the nuclear motion on the femtosecond time scale may control
the fragmentation patterns (charge-directed reactivity). Similar effects
should be present in neutral systems, with ionization replaced by
coherent excitation of several electronic states \cite{vibok1,vibok2,vibok3,vibok4,vacher1,vacher2,remacle2}.
While in ionization one talks about the dynamics of a hole \cite{santra}
and its coupling to nuclear degrees of freedom, in a neutral system
one can talk about the coupling between the exciton and the nuclei
\cite{kuleff1}. 

In the present work we consider the possibility of creating an electronic
wavepacket as a coherent superposition of two excited electronic states
in a neutral and rigid polyatomic molecule so as to get insight into
the lifetime of electronic coherence and to decipher to which extent
the nuclear motion may be detrimental. Pyrazine was chosen as a prototypical
system \cite{bolovinos,domcke,woywod1,woywod2,lenz4,woywod3,woywod4,guerin1,guerin2}
in the present work for several reasons. We studied in a series of
previous works \cite{vibok1,vibok2,vibok3,vibok4} the conditions for creating such an electronic wavepacket
in a neutral system, namely ozone, both theoretically and experimentally,
and observed long-lived electronic coherence (for some time after
the pump pulse was extinct), which we related to vibrational oscillations
of the fraction of the system that does not dissociate immediately
after light absorption. In contrast, we chose pyrazine here because
it is a bound system that does not dissociate when populating its
excited electronic states in the MUV/FUV (middle/far ultraviolet)
spectral domain ($\sim$ 5--7 eV here) by a few-femtosecond
laser pulse. Such a preparation resembles the initial conditions that
are the focus of attophysics with subfemtosecond pulses. The latter
are, at the moment, able to create electronic wavepackets in molecular
cations generated by photoionisation in the XUV (extreme ultraviolet).
Our perspective is different: we specifically want to stay below the
ionisation threshold so as to create an electronic wavepacket in a
neutral molecule in order to pave the way for attochemistry. In this
context, the subfemtosecond ($\sim$ 10--100 as) pulse would not serve
as a pump, but as a probe of the electronic motion.

The objective of the present study is to decipher the effect
of the nuclear motion on the preservation of electronic coherence,
more specifically with respect to two aspects: the anharmonicity of
the potential energy surfaces and the coordinate-dependence of the
transition dipole moments (TDMs). To this end, we define first an ideal model where
we make three approximations: (i) only the five totally symmetric
modes move (which, as a consequence, preserves symmetry and prevents nonadiabatic couplings
to act), (ii) they correspond to uncoupled harmonic oscillators, and
(iii) the TDMs from the ground electronic state to the
two bright states populated by the laser pulse are constant (Franck-Condon
approximation). We then lift the second and third approximations by
considering, first, the effect of anharmonicity, second, the effect
of coordinate-dependence of the TDMs (first-order Herzberg-Teller
contributions), third, both together.

We will show that such conditions can create an electronic wavepacket in a neutral molecule that induces oscillations of the electric dipole and quadrupole.
This may trigger new types of experimental observations of great relevance for the attophysics community.
We shall also examine the conditions for electronic coherence to survive on the longer term between the two excited
states and how it is correlated to the compacity of the nuclear density and its capability of reconstructing itself over time.

The structure of the paper is organized as follows. Section II. provides
details on the electronic structure and dynamical calculations,
as well as information about the model Hamiltonians. In section III., results for
the different models are presented and discussed, and
conclusions are given in section IV.


\begin{figure*}[ht]
\includegraphics[clip,width=17cm]{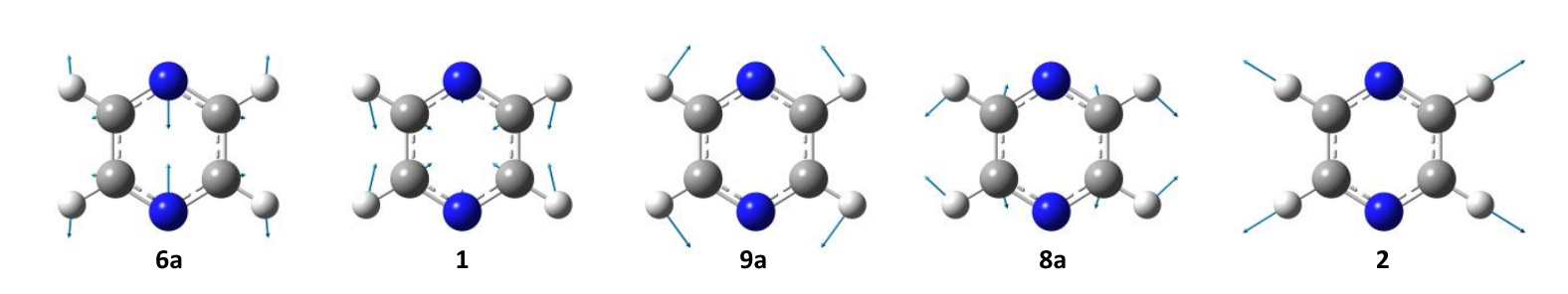} \caption{
The five totally symmetric vibrational modes of ground-state pyrazine obtained with DFT PBE0/AVTZ calculations (mode nomenclature known as Lord's system ~\cite{Lord57}, further specified in Ref. ~\cite{Innes88}; see below in the text). Mode 6a: ring bending; mode 1: ring breathing; mode 9a: CH rocking; mode 8a: ring stretching (quinoidal); mode 2: CH stretching; all are in-plane distortions. 
}
\label{fig:modes} 
\end{figure*}

\vspace{0.5cm}
\leftline {\bf II. METHODOLOGY}
\vspace{0.5cm}

\vspace{0.5cm}
\leftline{{II.1 MOLECULAR GEOMETRIES, VIBRATIONAL }}
\leftline{{MODES, ORBITALS, AND ELECTRONIC STATES }}
\leftline{{OF INTEREST}}
\vspace{0.5cm}

\vspace{0.5cm}
\leftline{\it{II.1.a Computational details - electronic structure }}
\vspace{0.5cm}

According to the literature (see for example Refs. ~\cite{woywod3,woywod4,Sala14,Kanno14} and references therein), MP2/AVTZ and CAS(10,8)PT2/AVTZ can be considered as state-of-the art levels of theory for the ground and valence excited states of pyrazine (yet with some open discussion remaining about the energies of some of the dark valence states and the performance of various flavours of MRPT including XMS-CAPST2 and XMCQDPT2 against MRCI, which are out of the scope of the present work). We used here DFT and TD-DFT out of convenience, as they provide various relevant properties for our purpose straightforwardly (see below). Our point here is not to reach \emph{ab initio} spectroscopic accuracy but to provide meaningful models to explore. 

As stated above, we performed electronic structure calculations with the DFT and TD-DFT approaches at the PBE0/AVTZ level of theory using the Gaussian 16, Revision A.03, quantum chemistry package ~\cite{Gaussian16A03}. We compared results given with this functional to those obtained with the CAM-B3LYP one. We also considered MP2/AVTZ results for comparison.
We give below a few points of comparison showing that our calculations seem meaningful enough for geometries and frequencies, yet require some slight adjustments in terms of vertical transition energies, so as to match experimental 0--0 band origins, as further explained in II.1.d.

First, let us state that the PBE0 exchange-correlation functional for DFT provides ground-state data that are close to MP2 quality and slightly better than CAM-B3LYP; see Table\ \ref{tab:geometry}. The ground-state equilibrium geometry is almost identical for all approaches with the AVTZ basis set. They agree with the experimental data provided in Ref.~\cite{Innes88} within less than 2$\%$ (root-mean-square deviation - RMSD - on the relative error: 1.6$\%$ for MP2/AVTZ, 1.7$\%$ for PBE0/AVTZ, and 1.9$\%$ for CAM-B3LYP/AVTZ).

 \begin{table}
 \footnotesize
 \caption{Ground-state equilibrium geometry: comparison between experimental data ~\cite{Innes88} and calculations (this work), with bond lengths given in \AA$\quad$and angles in degrees.}\label{tab:geometry}
\begin{tabular}{lcccc}\hline\hline
      & Exp. & \quad MP2/AVTZ\quad  &\quad CAM-B3LYP/AVTZ\quad  &\quad PBE0/AVTZ \\\hline\hline
 CC   &  1.403    &  1.393  &  1.386  &  1.389  \\
 CN   &  1.339    &  1.339  &  1.325  &  1.327  \\
 CH   &  1.115    &  1.083  &  1.083  &  1.085  \\
 NCC  &  122.2    &  122.3  &  121.8  &  122.0  \\
 CNC  &  115.6    &  115.4  &  116.4  &  116.0  \\
 NCH  &  113.9    &  116.9  &  117.3  &  117.2  \\\hline\hline
 \end{tabular}
 \end{table}

We also compared ground-state frequencies at the equilibrium geometry. Again, PBE0/AVTZ performs slightly better than CAM-B3LYP/AVTZ compared to both MP2/AVTZ calculations and experimental data ~\cite{Innes88}. Here, we report frequencies for the five totally symmetric modes that are considered in our following investigation; see Table\ \ref{tab:vibrations} and Fig.\ \ref{fig:modes}. Again, they are almost identical for all approaches. They agree with the experimental data provided in Ref. ~\cite{Innes88} within less than 5$\%$ (RMSD on the relative error: 2.7$\%$ for MP2/AVTZ, 3.5$\%$ for PBE0/AVTZ, and 4.2$\%$ for CAM-B3LYP/AVTZ). 

In all cases, the largest discrepancies between experimental and theoretical data in terms of ground-state geometries and vibrations concern the CH bonds. They have little impact regarding electronic structure in the present study, which is focussed on the $\pi$-system of pyrazine. Note that excited-state calculations will be further discussed below, in II.1.d.

 \begin{table}
 \footnotesize
 \caption{Ground-state vibrations: comparison between experimental data ~\cite{Innes88} and calculations (this work), with fundamental frequencies given in cm$^{-1}$ (mode nomenclature known as Lord's system ~\cite{Lord57}, further specified in Ref. ~\cite{Innes88}; see below in the text).}\label{tab:vibrations}
\begin{tabular}{lcccc}\hline\hline
     & Exp. & \quad MP2/AVTZ\quad  &\quad CAM-B3LYP/AVTZ\quad  &\quad PBE0/AVTZ \\\hline\hline
 6a  &  596   &  597   & 617  & 606   \\
 1   &  1015  &  1028  & 1065 & 1058   \\
 9a  &  1230  &  1250  & 1266 & 1257   \\
 8a  &  1582  &  1610  & 1656 & 1643   \\
 2   &  3055  &  3217  & 3197 & 3189  \\\hline\hline
 \end{tabular}
 \end{table}

\vspace{0.5cm}
\leftline{\it{II.1.b Symmetry considerations }}
\vspace{0.5cm}

\begin{figure}[ht]
\includegraphics[clip,width=6cm]{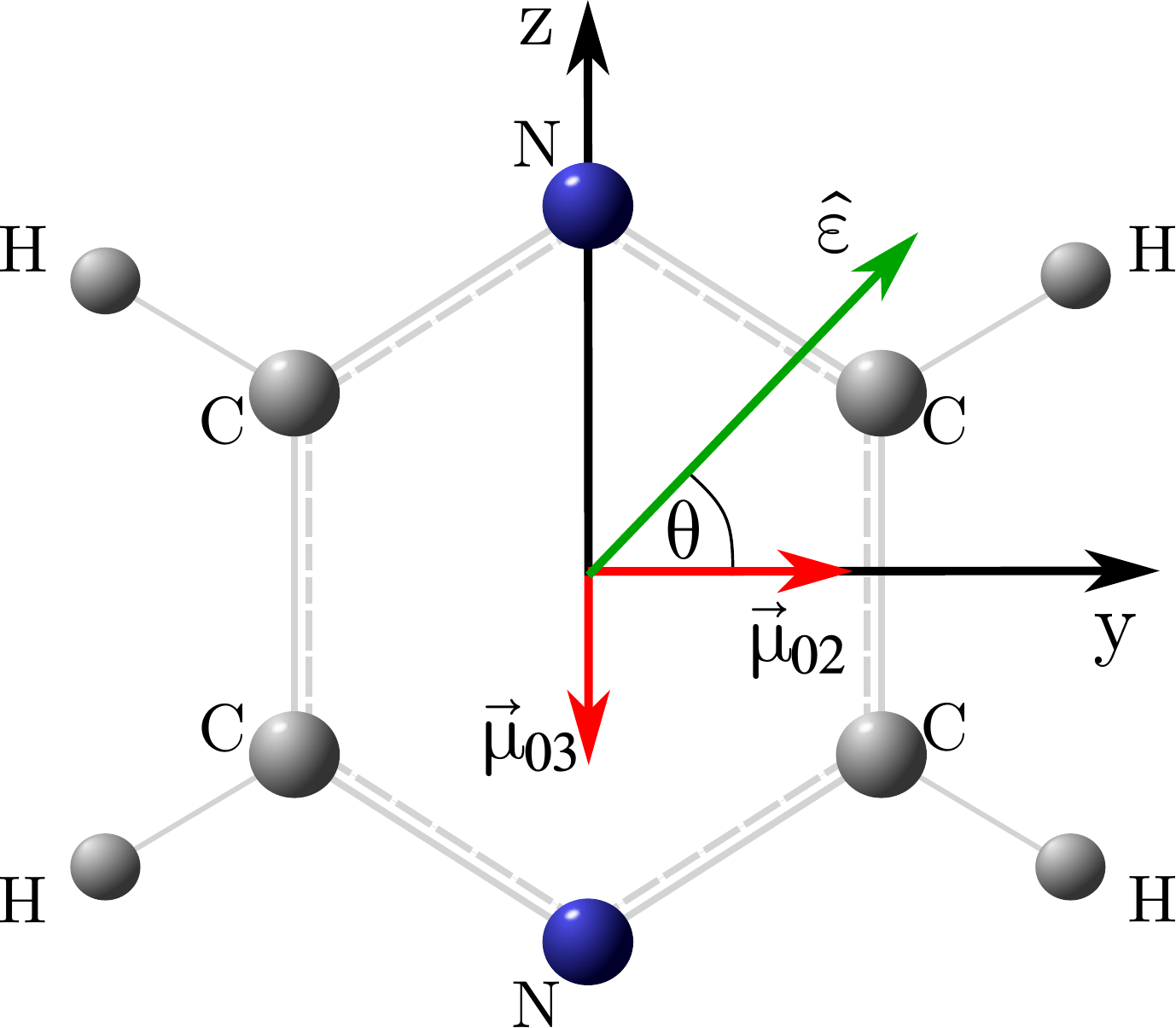} \caption{The body-fixed axes of the pyrazine molecule. 
The TDM vectors of the $S_2$ and $S_3$ excited states with the $S_0$ ground state are denoted by $\vec{\mu}_{02}$ and $\vec{\mu}_{03}$, respectively.
The laser polarization of the UV pump pulse ($\hat{\varepsilon}$) is set at $45^{\circ}$ with respect to the $+y$ and $+z$ directions.
}
\label{fig:system} 
\end{figure}

The equilibrium geometry of the pyrazine molecule (C$_4$N$_2$H$_4$; see Fig.\ \ref{fig:modes}) in its electronic ground-state ($S_0$ $1 ^{1}\mathrm{A_g}$) belongs to the $D_{2h}$ point group. In what follows, the body-fixed Cartesian frame, centred at the centre of mass of the molecule, will be considered as aligned with its principal axes of inertia (see Fig.\ \ref{fig:system}). For $D_{2h}$ geometries and using Mulliken's convention, $x$ is orthogonal to the molecular plane and $z$ goes through both N-atoms. Accordingly, $x$, $y$, and $z$ generate the $B_{3u}$, $B_{2u}$, and $B_{1u}$ irreducible representations of $D_{2h}$, respectively.

\vspace{0.5cm}
\leftline{\it{II.1.c Valence orbitals }}
\vspace{0.5cm}

\begin{figure}[ht]
\includegraphics[clip,width=8cm]{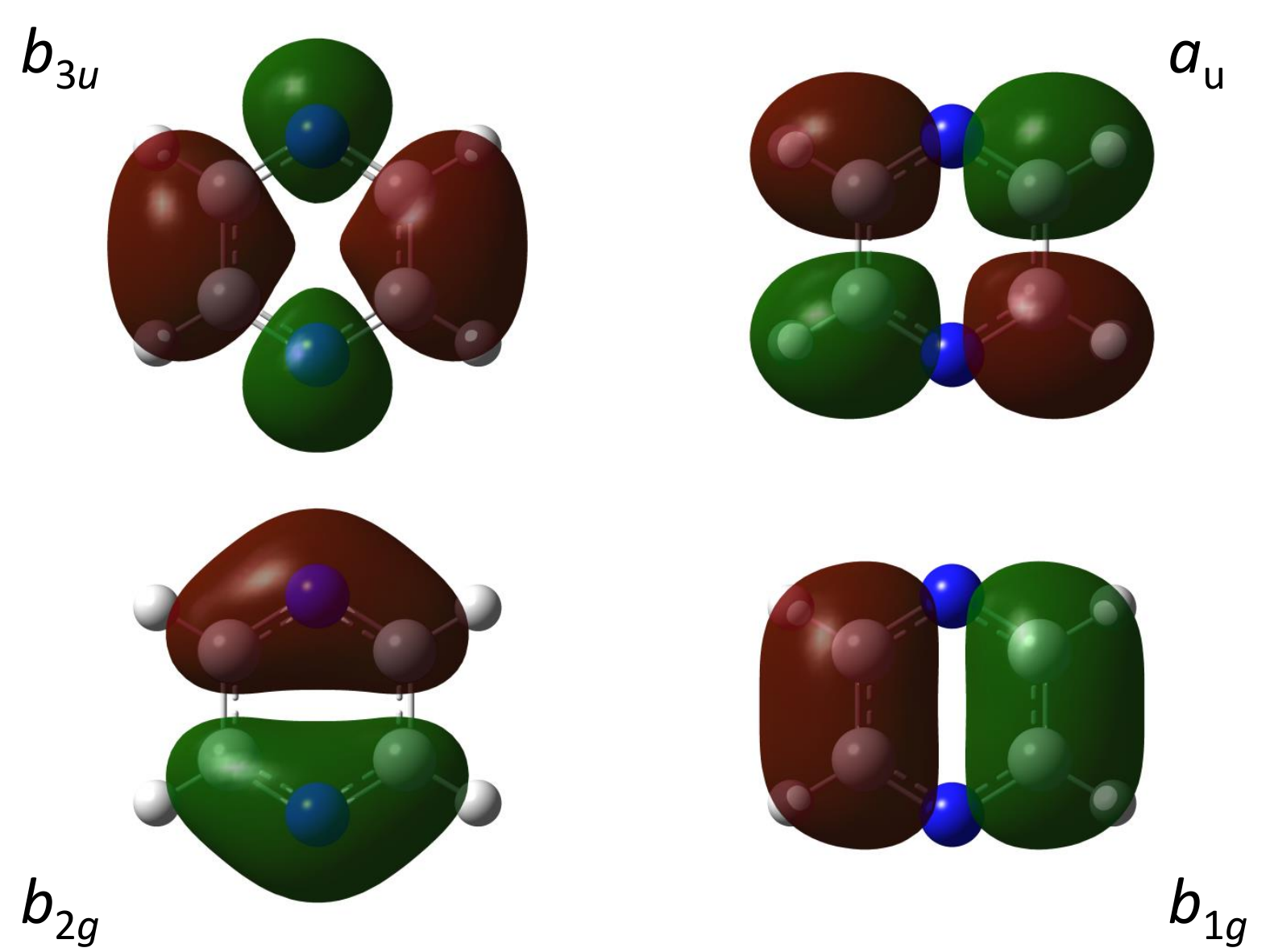} \caption{The four valence $\pi/\pi^{*}$-orbitals of pyrazine involved in the excited states discussed in the present work (all are antisymmetric with respect to the molecular plane). 
}
\label{fig:pipistar} 
\end{figure}

There are eight valence orbitals in pyrazine, containing ten valence electrons: a six-centre $\pi$-system (three bonding and three antibonding out-of-plane orbitals with six electrons describing the six benzene-like heteroaromatic bonds of the C$_4$N$_2$-ring) and a two-centre $n$-system (two nonbonding in-plane orbitals with four electrons describing the lone pairs on both N-atoms). This typically calls for a minimal complete active space CAS(10,8) in order to account for static electron correlation (see for example Refs. ~\cite{woywod3,woywod4,Sala14,Kanno14} and references therein). Among them, the most relevant orbitals for describing the low-lying valence electronic states of pyrazine in terms of dominant one-electron transitions are: $b_{2g}$ ($\pi$, $zx$-type), $b_{1u}$ ($n$, antisymmetric), $b_{1g}$ ($\pi$, $yx$-type), $a_g$ ($n$, symmetric), $b_{3u}$ ($\pi^{*}$, $(z^2-y^2)x$-type), and $a_u$ ($\pi^{*}$, $(2zy)x$-type); see Fig.\ \ref{fig:pipistar} for the $\pi/\pi^{*}$-orbitals. The next virtual orbital of $a_g$ symmetry, which is known to have a mixed Rydberg/$\sigma^{*}$(CH) character according to, \emph{e.g.}, Ref. ~\cite{Stener11}, is involved in the lowest-lying Rydberg-type states, in agreement with our calculations.

\vspace{0.5cm}
\leftline{\it{II.1.d Electronic states }}
\vspace{0.5cm}

As shown in Ref. \cite{bolovinos}, which provides an extensive description of the absorption spectra of benzene and several azabenzenes in the UV domain, the excited singlet electronic states of pyrazine are very similar to those of benzene with respect to $\pi \rightarrow \pi^{*}$ transitions, except that optical selection rules in the $D_{2h}$ point group of pyrazine are less strict than in the $D_{6h}$ point group of benzene, thus making all $\pi \rightarrow \pi^{*}$ states bright here. Also, azabenzenes involve extra $n \rightarrow \pi^{*}$ transitions that yield supplementary valence states and absorption bands. In the present work, we shall focus on the two lowest-lying $\pi \rightarrow \pi^{*}$ states. 

The two excited singlet electronic states of pyrazine that probably are the most famous in the literature are usually denoted  $S_1$ ($1 ^{1}\mathrm{B_{3u}}$; $n \rightarrow \pi^{*}$ character), around 4 eV, and $S_2$ ($1 ^{1}\mathrm{B_{2u}}$; $\pi \rightarrow \pi^{*}$ character), around 5 eV. Both are valence states and symmetry allowed, but the first one is almost dark (weak wavefunction overlap with the ground state), while the second one is significantly bright (locally excited state). They are known to be nonadiabatically interacting together through strong vibronic couplings around a conical intersection that lies close to the Franck-Condon point, involving a single vibrational mode (sole $B_{1g}$ mode known as 10a, which essentially describes the relative torsion of the H$_4$-plane with respect to the C$_4$N$_2$-ring around the N$_2$-axis). This coupling results on abnormal broadening and loss of vibrational structure in the absorption spectrum of the $S_2$ state. Extensive studies of this effect have made pyrazine one of the most famous prototypical benchmarks for nonadiabatic quantum dynamics simulations, for example with the multiconfiguration time-dependent Hartree (MCTDH) method ~\cite{lenz4,Sala14}, its multilayer extension ML-MCTDH ~\cite{Vendrell11}, and the variational multiconfiguration Gaussian (vMCG) method ~\cite{Penfold19}.

Hereafter, we shall focus exclusively on the aforementioned $S_2$ ($1 ^{1}\mathrm{B_{2u}}$; $\pi \rightarrow \pi^{*}$ character) state together with the next bright valence state that will be denoted here $S_3$ ($1 ^{1}\mathrm{B_{1u}}$; $\pi \rightarrow \pi^{*}$ character). The UV one-photon absorption spectrum of pyrazine \cite{bolovinos} shows their spectral signature as two close but nonoverlapping bands of similar intensities, centred around 5 and 6.5 eV, respectively. They are characterised by two in-plane TDMs (along $y$ and $z$, respectively). The corresponding oscillator strengths are estimated at 6 and 10, respectively, according to experimental data \cite{bolovinos}. We chose this pair of states for their similar oscillator strengths and because their spectral range makes it possible for them to be populated together and coherently by the same pulse with a spectral width of about 2.6 eV around 6 eV corresponding to a typical duration of 1 fs. In addition, they are below the onset of the Rydberg states (\emph{ca.} 7--8 eV). Hence, the $S_2$ and $S_3$ states of pyrazine are excellent candidates for creating an electronic wavepacket made of a superposition of two states in a neutral organic molecule of decent size. We thus expect the present work to trigger realistic experiments in this context. 

In the present work, we deliberately consider early dynamics along the five vibrational modes that preserve the $D_{2h}$ point group of ground-state pyrazine. The totally-symmetric $A_g$ modes are labelled 6a, 1, 9a, 8a, and 2 ~\cite{Innes88}; see Fig.\ \ref{fig:modes}. They are the only five distortions along which there can be an energy gradient in the Franck-Condon regions of the excited electronic states. As such, we expect finite variations of $\langle q \rangle$ (mean position) and $\langle p \rangle$ (mean momentum) along them only. From a quasiclassical perspective, this means that group velocity will only develop along them, inducing wavepacket displacement, while phase velocity will modify the other modes later on, once intramolecular vibrational redistribution starts taking place, thus inducing wavepacket spreading around zero mean position and momentum. As we are here interested in the early stages of the photodynamics of pyrazine, it is legitimate to consider the five $A_g$ modes as prominent. In terms of spectroscopy, they are the only modes susceptible of inducing vibrational progressions in the absorption spectra within the Franck-Condon approximation, as will be illustrated later on.

Within such a perspective, we looked for the lowest-lying stationary points in $S_2$ and $S_3$ where $D_{2h}$ symmetry stays preserved. As it occurs, the apparent minimum in $S_2$ is a saddle point of index 2 (unstable along $B_{2g}$ mode 4, chair-type ring puckering, and $B_{3u}$ mode 16b, boat-type ring puckering) and that in $S_3$ is a saddle point of index 1 or transition state (unstable along $A_u$ mode 16a, twist-type ring puckering). The geometries of these two points, the frequencies of their five $A_g$ modes, as well as relevant transition energies, are provided in Table\ \ref{tab:excited} for the level of theory used in our quantum chemistry calculations.

 \begin{table}
 \footnotesize
 \caption{Excited-state properties calculated at the TD-DFT PBE0/AVTZ level of theory.
 Bond lengths are given in \AA$\quad$and angles in degrees.
 $A_g$-mode fundamental frequencies are given in cm$^{-1}$.
 Energies ($E_s$) are given in eV for each state $s$, with corresponding oscillator strengths ($f_s$) from the ground state to excited states $s$.
 The energy zero corresponds to $-264.098160$ hartree at the ground-state ($1 ^{1}\mathrm{A_g}$) minimum.
 The excited apparent ``minima'' are saddle points of index 2 and 1 for the $1 ^{1}\mathrm{B_{2u}}$ and $1 ^{1}\mathrm{B_{1u}}$ states, respectively.
 Band origins ($0-0$) given in eV are only estimates within the five-dimensional subspace and with respect to such saddle points.
 Ground-state properties are recalled for easier comparison (see Tables\ \ref{tab:geometry} and Table\ \ref{tab:vibrations}).
 }\label{tab:excited}
\begin{tabular}{lccc}\hline\hline
                & $A_g$ min & $B_{2u}$ ``min'' & $B_{1u}$ ``min'' \\\hline\hline
 CC             & 1.389     & 1.422            & 1.512 \\                                   
 CN             & 1.327     & 1.358            & 1.320 \\                                
 CH             & 1.085     & 1.086            & 1.084 \\                                
 NCC            & 122.0     & 125.1            & 121.4 \\
 CNC            & 116.0     & 109.7            & 117.2 \\
 NCH            & 117.2     & 116.8            & 118.6 \\\hline
 6a             &  606      &  539             &  565  \\
 1              & 1058      &  991             &  899  \\
 9a             & 1257      & 1245             & 1219  \\
 8a             & 1643      & 1562             & 1438  \\
 2              & 3189      & 3188             & 3202  \\\hline
 $E_{A_g}$      & 0.00      & 0.25             & 0.54  \\
 $E_{B_{2u}}$   & 5.51      & 5.28             & 5.73  \\
 $E_{B_{1u}}$   & 6.60      & 6.56             & 6.20  \\\hline
 $0-0_{B_{2u}}$ &           & 5.27             &       \\
 $0-0_{B_{1u}}$ &           &                  & 6.17  \\\hline
 $f_{B_{2u}}$   & 0.11      & 0.11             & 0.12  \\
 $f_{B_{1u}}$   & 0.07      & 0.07             & 0.16  \\\hline\hline
\end{tabular}
\end{table}

\vspace{0.5cm}
\leftline{{II.2 THREE-STATE FIVE-MODE MODELS }}
\vspace{0.5cm}

We consider here four models with increasing complexity: H+FC (harmonic and Franck-Condon), H+HT (harmonic and Herzberg-Teller to first order), A+FC (anharmonic and Franck-Condon), and A+HT (anharmonic and Herzberg-Teller to first order). In all cases, the five nuclear degrees of freedom are the mass-weighted normal coordinates along the five $A_g$ modes of ground-state pyrazine (see Fig.\ \ref{fig:modes}). The three electronic states are $S_0$, $S_2$, and $S_3$ (see Subsec. II.1.d). Within the $D_{2h}$ subspace, there are no nonadiabatic couplings among them. They are coupled only via light-matter interaction: $S_0$ with $S_2$ along $y$ ($B_{2u}$ TDM, $\vec{\mu}_{02}$) and $S_0$ with $S_3$ along $z$ ($B_{1u}$ TDM, $\vec{\mu}_{03}$).

In the simplest model, H+FC, the three adiabatic Hamiltonian operators are uncoupled and are harmonic and centred at the minimum of $S_0$ or at the apparent minima of $S_2$ and $S_3$, respectively (see Table\ \ref{tab:excited}). They were obtained upon projecting the corresponding Hessian matrices onto the five-dimensional $A_g$ subspace (accounting for Duschinsky rotation matrices and shift vectors). 
The corresponding 0--0 band origins calculated within the reduced five-dimensional subspace are at 5.27 and 6.17 eV. They are in the correct spectral region for our purpose, but their interpretation is delicate, since they correspond here to saddle points rather than actual minima.
The energies of the $S_2$ and $S_3$ apparent minima were thus further shifted with respect to the energy of the $S_0$ minimum so as to match the experimental 0--0 band origins at 4.69 eV and 6.30 eV, respectively \cite{bolovinos}. We considered the values of the TDMs or related oscillator strengths (see Table\ \ref{tab:excited}) at the $S_0$ origin (Franck-Condon point). The anharmonic models were obtained upon replacing the quadratic functions along each mode by Morse functions with identical curvatures at the minima and dissociation energies obtained by a fitting procedure over a range of representative geometries made of about 50 geometry-energy points per electronic state. The Herzberg-Teller models were generated by adding the first-order expansion of both TDMs along all coordinates from the Franck-Condon point (linear Herzberg-Teller approximation). The relevant mathematical expressions and parameter values of the various models are provided in SI.

\begin{figure}[ht]
\includegraphics[clip,width=8.5cm]{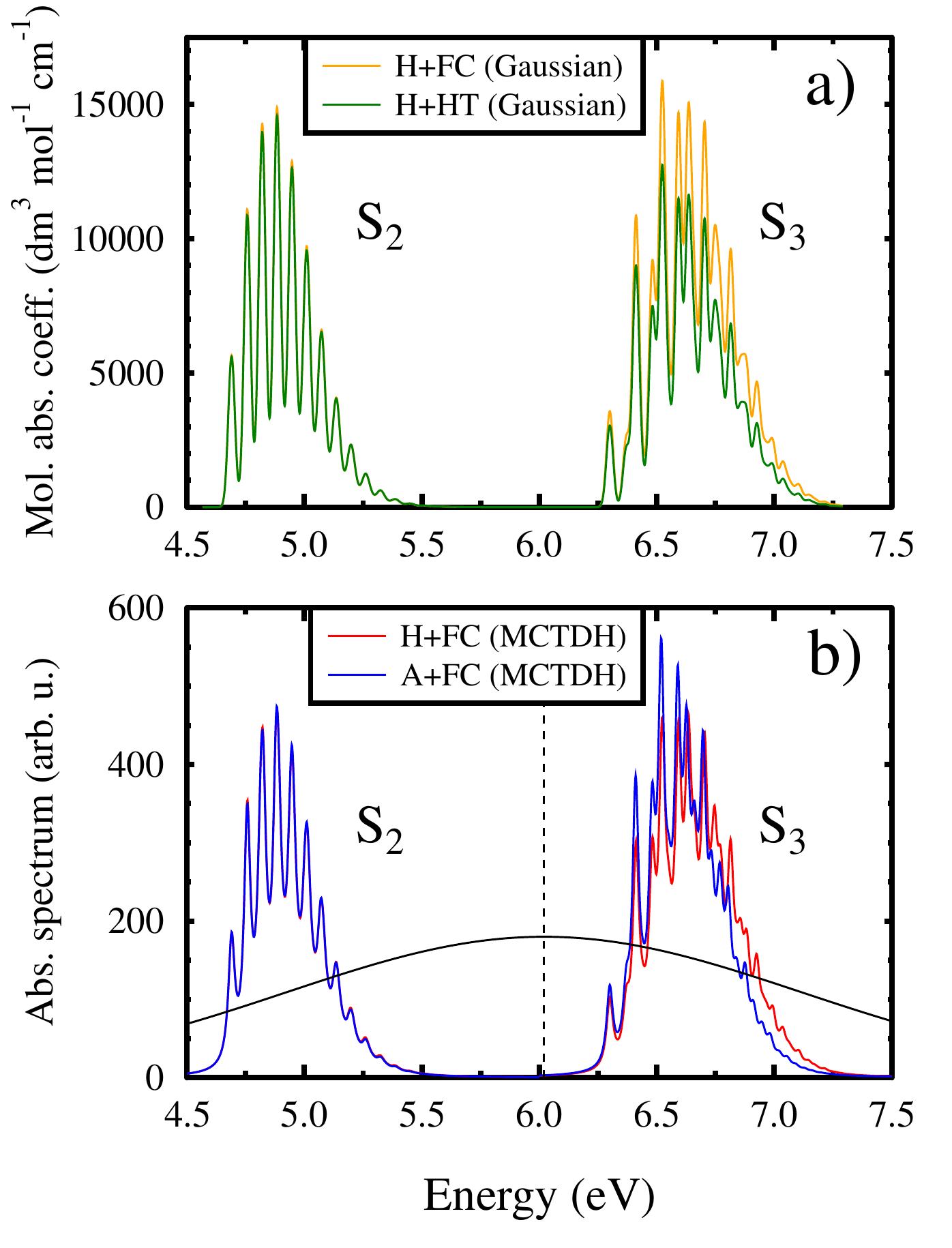} \caption{
(a) Absorption spectra of the $S_2$ and $S_3$ electronic states of the H+FC and H+HT pyrazine models calculated with GAUSSIAN.
(b) Absorption spectra of the $S_2$ and $S_3$ electronic states of the harmonic and anharmonic pyrazine models calculated through eq.\ \ref{eq:Sp} ($\tau = 40$ fs) using MCTDH.
The vertical dashed line indicates the central energy ($\hbar\omega_p = 6.02$ eV) while the solid black line shows the spectrum of the pump pulse applied in the present work. 
}
\label{fig:spectra2} 
\end{figure}

In addition to our time-resolved investigation of coherences and populations induced by an explicit laser pulse, we first compared the effect of the various
types of approximations on the absorption spectra. Vibrationnally resolved absorption spectra (so-called vibronic spectra) ~\cite{Santoro08} were calculated via the FCHT module implemented within the quantum chemistry package Gaussian 16, Revision A.03 ~\cite{Gaussian16A03}, according to two frameworks: the H+FC and H+HT models. According to our dimensionality reduction scheme, we limited the set of vibronic transitions from the ground state to one or the other excited state to the subspace of the five $A_g$ modes, which are those expected to provide dominant vibrational progressions. Within this procedure, the intensities of the peaks were obtained from explicit formulae for the Franck-Condon factors (and also first-order Gaussian moments with the Herzberg-Teller picture) between two multidimensional linear harmonic oscillators with displaced minima (shift vectors) and rotated modes (Duschinksy matrices). The stick spectra were further convoluted with band broadening simulated by means of Gaussian functions with half-widths at half-maximum of 135 cm$^{-1}$ (default value). As shown in  Fig.\ \ref{fig:spectra2}, first-order Herzberg-Teller contributions have negligible effect on the first spectrum,
($^{1}\mathrm{A_{g}}$ $\rightarrow$ $^{1}\mathrm{B_{2u}}$), but significant one ($\sim$ 20$\%$) on the second spectrum ($^{1}\mathrm{A_{g}}$ $\rightarrow$ $^{1}\mathrm{B_{1u}}$). 

We also calculated absorption spectra with the quantum dynamics package MCTDH 84.10 ~\cite{mctdh1,mctdh2} from Fourier-transformed autocorrelation functions, now according to the H+FC and A+FC models. The H+FC spectra obtained from the two methods (Franck-Condon factors or wavepacket propagation) are almost identical (within numerics, according to different broadening procedures), as expected. For the wavepacket approach, the spectral broadening was simulated in the time domain using a damping time of 40 fs. The A+FC spectrum obtained with the time-dependent method is very similar to the H+FC one, with only small differences of the peak positions in the high-energy part, consistent with the fact that the system is bound and quite rigid (see Fig.\ \ref{fig:spectra2}). 

\vspace{0.5cm}
\leftline{{II.3 NUCLEAR DYNAMICS}}
\vspace{0.5cm}

\vspace{0.5cm}
\leftline{\it{II.3.a The nuclear Hamiltonian}}
\vspace{0.5cm}

In the Hilbert subspace of the three electronic states, the general form of the time-dependent nuclear Hamiltonian that incorporates the laser-molecule interaction reads
\begin{equation}\label{eq:full}
\hat{H}(t) = \begin{pmatrix} \hat{H}_{0} & - \vec{\mu}_{02} ({\bf Q}) \cdot \vec{E}(t)  & - \vec{\mu}_{03} ({\bf Q}) \cdot \vec{E}(t) \\ - \vec{\mu}_{02} ({\bf Q}) \cdot \vec{E}(t) & \hat{H}_{2}  & 0  \\ -\vec{\mu}_{03} ({\bf Q}) \cdot \vec{E}(t) & 0 & \hat{H}_{3} \end{pmatrix}  \quad\quad ,
\end{equation}
where $\hat{H}_{\alpha} = \hat{T} + \hat{V}_{\alpha}({\bf{Q}})$ corresponds to the $\alpha$th electronic state ($\alpha=0, 2, 3$),  
$\hat{T}=-\frac{1}{2}{\bf{\nabla_Q^{\dagger} \nabla_Q }}$ is the kinetic energy operator (assuming mass-weighted coordinates and atomic units with $\hbar=1$ a.u.) and $\hat{V}_{\alpha}({\bf{Q}})$ is the potential energy function, which is considered 
either harmonic or anharmonic in this work. For the explicit form of the harmonic and anharmonic potentials and corresponding parameter values, see SI.
In eq.\ \ref{eq:full}, the light-molecule interaction is considered within the electric dipole approximation with 
$\vec{\mu}_{0\alpha}$ being the TDM of the $\alpha$th electronic state with the ground state and $\vec{E}(t)$ a linearly polarized Gaussian electric field 
\begin{equation}\label{eq:pump}
\vec{E}(t) = E_0 g(t) \cos(\omega_p (t-t_0)) \vec{\varepsilon} .
\end{equation}
In eq.\ \ref{eq:pump}, $E_0$ is the electric field amplitude, $\vec{\varepsilon}$ is the polarization unit vector, $\omega_p$ is the central angular frequency, $t_0$ is the center of the pulse, and $g(t)=\exp(-(t-t_0)^2/2\sigma^2)$ is a Gaussian envelope function with $t_p=2 \sigma \sqrt{\ln(2)}$ being the pulse typical duration.

\vspace{0.5cm}
\leftline{\it{II.3.b Computational details - quantum dynamics}}
\vspace{0.5cm}

The time-dependent nuclear Schr\"odinger-equation was solved in the present work with the multiconfiguration time-dependent Hartree (MCTDH) method~\cite{mctdh1,mctdh2}. The wavefunction representation of MCTDH is based on time-independent primitive basis sets ($\chi$) that are then used to construct single-particle functions ($\phi$) whose time-dependent linear combinations form the total nuclear wavepacket ($\psi$),
\begin{equation}\label{eq:M1}
\phi^{(k)}_{j_{k}}(Q_k,t)=\sum\limits_{i=1}^{N_{k}} c^{(k)}_{j_{k}i}(t)\chi^{(k)}_{i}(Q_k) ,   \quad \quad \quad  k=1, 2, 3, 4, 5 
\end{equation}
\begin{equation}\label{eq:M2}
\psi({\bf Q}, t)=\sum\limits_{j_{1}=1}^{n_{1}} ...  \sum\limits_{j_{5}=1}^{n_{5}} A_{{j_1} ... {j_5}}(t)  \prod_{k=1}^5 \phi^{(k)}_{j_{k}}(Q_k,t)  .
\end{equation}
In our calculations, Hermite functions (eigenfunctions of linear harmonic oscillators) were taken as primitive basis functions along each coordinate, the number of which were ranging from $N_k = 7, ..., 57$ depending on the $Q_k$ coordinate. 
The number of single particle functions were ranging from $n_{k} = 4, ..., 8$. With these parameter values, correct convergence of the dynamical calculations has been ensured.

\vspace{0.5cm}
\leftline{\it{II.3.c Calculated dynamical quantities}}
\vspace{0.5cm}

Utilizing the nuclear wavepacket (see previous sub-section) with coupled components on each of the three electronic states, several dynamical quantities were calculated to reveal the temporal behaviour of the system under consideration.
The electronic state populations are obtained from the formula
\begin{equation}\label{eq:Po}
P_i (t) = \langle \psi^{(i)} ({\bf Q}, t) | \psi^{(i)} ({\bf Q}, t) \rangle ,         
\end{equation}
where $\psi^{(i)} ({\bf Q}, t)$ is the projection of the total wavepacket $\psi({\bf Q},t)$ on the considered electronic state ($i=0, 2, 3$) and integration is performed over ${\bf Q}$-space.

The electronic coherence between the $S_2$ and $S_3$ states is calculated according to the reduced overlap integral,
\begin{equation}\label{eq:Co}
C(t) = \frac{\langle \psi^{(2)} ({\bf Q}, t) | \psi^{(3)} ({\bf Q}, t) \rangle} {\sqrt{P_{2}(t) P_{3}(t)}} .     
\end{equation}
The compacity, which measures how much the system is expanded during the time evolution, is calculated as the expectation value of the sum of the coordinate squares,
\begin{equation}\label{eq:Cp}
K(t) = \langle \sum_{i=1}^5 Q_i^2 \rangle  .    
\end{equation}
The autocorrelation function is determined from the following formula
\begin{equation}\label{eq:Ac}
A(t) =  \langle \psi (t=0) | \psi (t) \rangle ,     
\end{equation}
where $\psi (t)$ is the total wavepacket and $\psi (t=0)$ is the initial wavefunction (nuclear ground state in the electronic ground state).
A so-called restricted autocorrelation function has been defined and calculated as
\begin{equation}\label{eq:Ar}
A^{\mathrm{(R)}}(t) =  \langle \psi_\mathrm{ref} | \psi (t) \rangle ,     
\end{equation}
where the reference function $\psi_\mathrm{ref}$ is prepared by a sudden excitation of the original initial state $\psi (t=0)$ half to $S_2$ and half to $S_3$ (nothing remaining on $S_0$).
In this way, as will be demonstrated in sec.III, $A^{(R)}(t)$ gets more correlated with the coherence $C(t)$ than the usual autocorrelation function $A(t)$, the time dependence of which being ``too much dominated'' by the largest component that remains in $S_0$.
Finally, the spectra of the excited electronic states was determined as the Fourier transform of the autocorrelation function, weighted by a usual damping function, and multiplied by the frequency (in arbitrary units)
\begin{equation}\label{eq:Sp}
S(\omega) =  \omega \int_{-\infty}^{\infty} A^{\mathrm{(R)}}(t) e^{-t/\tau} e^{i \omega t} dt  ,
\end{equation}
where $\tau$ is a typical time damping parameter chosen so as to remove the numerical boundary effects of finite-time calculations.


\vspace{0.5cm}
\leftline {\bf III. RESULTS AND DISCUSSION}
\vspace{0.5cm}

\vspace{0.5cm}
\leftline{{III.1 ELECTRONIC WAVEPACKET }}
\vspace{0.5cm}

Pyrazine contains 42 electrons. What we call ``core'' hereafter is made of 38 electrons in 19 closed orbitals (inner core shells, valence $\sigma$-bonds together with the lowest $\pi$-bond, and lone pairs on nitrogen atoms), which will be distinguished from the frontier $\pi$-system, made of four active electrons in four active orbitals. The $S_0$ state, $^{1}\mathrm{A_{g}}$, is dominated by the closed-shell configuration (represented by a single Slater Kohn-Sham determinant), 
\begin{equation}\label{eq:S0}
^{1}\mathrm{\Phi_{0}} = {^{1}\mathrm{\Phi_{A_{g}}}} = | (\mathrm{core}) b_{2g} \overline{b_{2g}} b_{1g} \overline{b_{1g}} |
\end{equation}
while the $S_2$ and $S_3$ states, $^{1}\mathrm{B_{2u}}$ and $^{1}\mathrm{B_{1u}}$, are dominated by linear combinations of two singly excited configurations. Within the Tamm-Dancoff approximation (TDA), they would read 
\begin{equation}\label{eq:S2}
^{1}\mathrm{\Phi_{2}} = {^{1}\mathrm{\Phi_{B_{2u}}}} = C_{b_{1g}}^{b_{3u}}  {^{1}\mathrm{\Phi_{b_{1g}}^{b_{3u}} }}  +  C_{b_{2g}}^{a_{u}}  {^{1}\mathrm{\Phi_{b_{2g}}^{a_{u}} }} , 
\end{equation}
\begin{equation}\label{eq:S3}
^{1}\mathrm{\Phi_{3}} = {^{1}\mathrm{\Phi_{B_{1u}}}} = C_{b_{1g}}^{a_{u}}  {^{1}\mathrm{\Phi_{b_{1g}}^{a_{u}} }}  +  C_{b_{2g}}^{b_{3u}}  {^{1}\mathrm{\Phi_{b_{2g}}^{b_{3u}} }} .
\end{equation}
Note that the TDA picture (CIS-like) is invoked here only for interpretation purposes but was not used in actual calculations. In other words, it occurs that the $X$-coefficients (excitations) of Cassida's TD-DFT equations dominate the $Y$-coefficients (deexcitations) by far here, but the numerical effect of the latter was not discarded from our calculations. 

In addition to the Slater Kohn-Sham determinant for the ground state, we consider hereafter four singly excited configurations, represented by four singlet configuration-state functions (CSFs), \emph{i.e}., spin-eigenfunctions of both ${\hat{S}}^2$ and ${\hat{S}}_z$. Each is a spin-symmetrised linear combination of two Slater determinants, 
\begin{equation}\label{eq:SS1}
^{1}\mathrm{\Phi_{b_{1g}}^{b_{3u}}} = \frac{1}{\sqrt{2}} \left ( | (\mathrm{core}) b_{2g} \overline{b_{2g}} b_{3u} \overline{b_{1g}} | + | (\mathrm{core}) b_{2g} \overline{b_{2g}} b_{1g} \overline{b_{3u}} | \right ) ,
\end{equation}
\begin{equation}\label{eq:SS2}
^{1}\mathrm{\Phi_{b_{2g}}^{a_{u}}} = \frac{1}{\sqrt{2}} \left ( | (\mathrm{core}) a_{u} \overline{b_{2g}} b_{1g} \overline{b_{1g}} | + | (\mathrm{core}) b_{2g} \overline{a_{u}} b_{1g} \overline{b_{1g}} | \right ) ,
\end{equation}
\begin{equation}\label{eq:SS3}
^{1}\mathrm{\Phi_{b_{1g}}^{a_{u}}} = \frac{1}{\sqrt{2}} \left ( | (\mathrm{core}) b_{2g} \overline{b_{2g}} a_{u} \overline{b_{1g}} | + | (\mathrm{core}) b_{2g} \overline{b_{2g}} b_{1g} \overline{a_{u}} | \right ) ,
\end{equation}
\begin{equation}\label{eq:SS4}
^{1}\mathrm{\Phi_{b_{2g}}^{b_{3u}}} = \frac{1}{\sqrt{2}} \left ( | (\mathrm{core}) b_{3u} \overline{b_{2g}} b_{1g} \overline{b_{1g}} | + | (\mathrm{core}) b_{2g} \overline{b_{3u}} b_{1g} \overline{b_{1g}} | \right ) .
\end{equation}
At the FC point, we have: $C_{b_{1g}}^{b_{3u}} = 0.9442 (89\%)$ and $C_{b_{2g}}^{a_{u}} = -0.3234 (10\%)$ [total: 100$\%$ with rounding];
$C_{b_{1g}}^{a_{u}} = 0.8872 (79\%)$ and $C_{b_{2g}}^{b_{3u}} = 0.4343 (19\%)$ [total: 98$\%$ with rounding].

\begin{figure}[ht]
\includegraphics[clip,width=8.5cm]{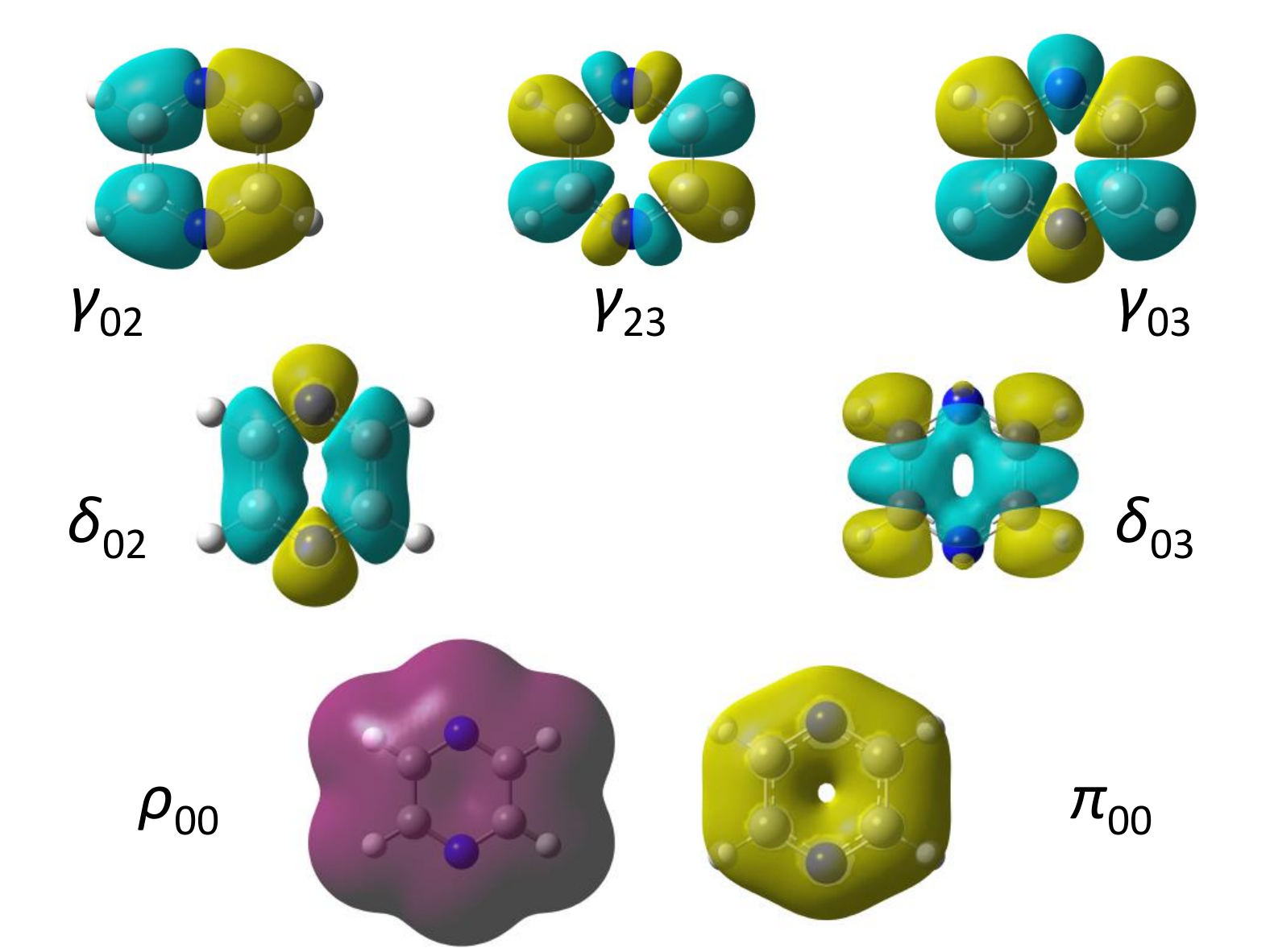} \caption{
The total, differential, and transition electron densities at the FC point. The total ground-state density (in purple), $\rho_{00}$, integrates to 42 electrons. The $\pi$-contribution, $\pi_{00}=\rho_{00} - \rho_{\mathrm{core}}=2[b_{2g}]^2 + 2[b_{1g}]^2$, integrates to four electrons. The negative charge excess (electron) is shown in dark yellow, the negative charge depletion (hole) in light blue, for differential and transition one-electron densities. 
}
\label{fig:transdens} 
\end{figure}

\begin{figure}[ht]
\includegraphics[clip,width=8.8cm]{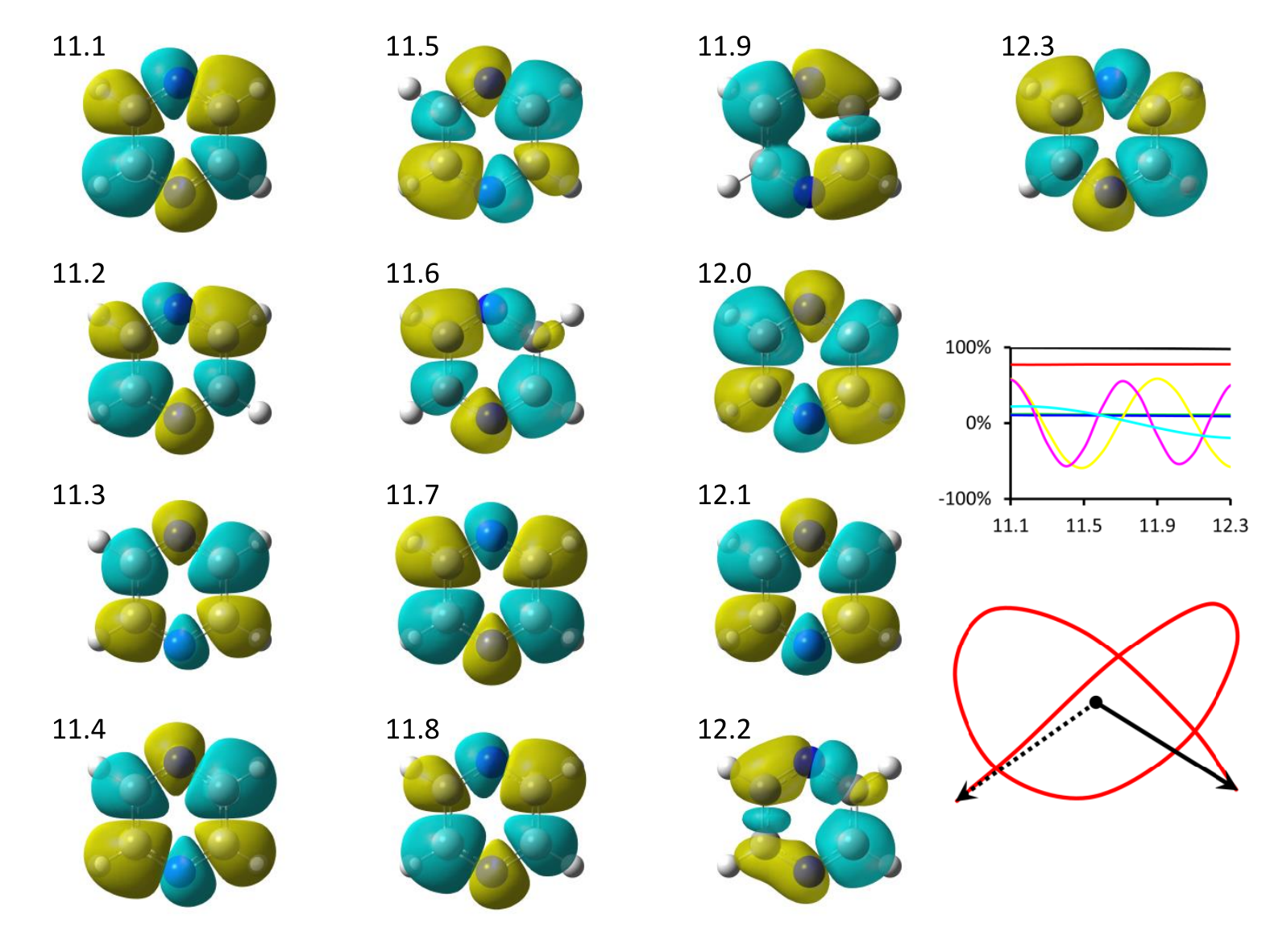} \caption{
The electronic wavepacket at the FC point. Electron excess is shown in dark yellow (negative charge: electron), electron depletion in light blue (positive charge: hole). The time evolution of the dipole moment is traced with a red curve (dashed arrow: at 11.1 fs; plain arrow: at 12.2 fs). Relative local populations and coherences against time are displayed in inset: 00 (red); 22 (green); 33 (blue); 02 (yellow); 03 (magenta); 23 (cyan); total (black). 
}
\label{fig:elecwp} 
\end{figure}

Canonical Kohn-Sham orbitals are natural orbitals for the ground-state determinant. The corresponding one-electron charge density thus reads as a sum of squared doubly occupied orbitals, 
\begin{equation}\label{eq:ro00}
\rho_{00}(\vec{r};{{\bf Q}_0}) = \rho_\mathrm{core}(\vec{r};{{\bf Q}_0}) + 2[b_{2g}(\vec{r};{{\bf Q}_0})]^2 + 2[b_{1g}(\vec{r};{{\bf Q}_0})]^2 .
\end{equation}
It is the probability density of finding any electron among 42 at position $\vec{r}$, whatever its spin state, and thus integrates to 42 (see Fig.\ \ref{fig:transdens}). The remainder, $\pi_{00}(\vec{r};{{\bf Q}_0})=2[b_{2g}(\vec{r};{{\bf Q}_0})]^2 + 2[b_{1g}(\vec{r};{{\bf Q}_0})]^2$, represents the density of the frontier $\pi$-system only and integrates to four. We then introduce, for better contrast, differential densities (excited \emph{vs.} ground state), $\delta_{02}(\vec{r};{{\bf Q}_0})$ and $\delta_{03}(\vec{r};{{\bf Q}_0})$, such that the corresponding one-electron charge densities of the excited states read 
\begin{equation}\label{eq:ro22}
\rho_{22}(\vec{r};{{\bf Q}_0}) = \rho_{00}(\vec{r};{{\bf Q}_0}) + \delta_{02}(\vec{r};{{\bf Q}_0}) ,
\end{equation}
\begin{equation}\label{eq:ro33}
\rho_{33}(\vec{r};{{\bf Q}_0}) = \rho_{00}(\vec{r};{{\bf Q}_0}) + \delta_{03}(\vec{r};{{\bf Q}_0}) .
\end{equation}
In each CSF, there is no transition density between the two Slater determinants involved together (they are related via double excitations). The differential densities thus have no cross-term and read 
\begin{equation}\label{eq:delta02}
\begin{split}
\delta_{02}(\vec{r};{{\bf Q}_0}) = \left [ C_{b_{1g}}^{b_{3u}} ({{\bf Q}_0}) \right ]^2  \left ( [b_{3u}(\vec{r};{{\bf Q}_0})]^2 - [b_{1g}(\vec{r};{{\bf Q}_0})]^2  \right ) \\
+ \left [ C_{b_{2g}}^{a_{u}} ({{\bf Q}_0}) \right ]^2 \left ( [a_{u}(\vec{r};{{\bf Q}_0})]^2 - [b_{2g}(\vec{r};{{\bf Q}_0})]^2  \right ) ,
\end{split}
\end{equation}

\begin{equation}\label{eq:delta03}
\begin{split}
\delta_{03}(\vec{r};{{\bf Q}_0}) = \left [ C_{b_{1g}}^{a_{u}} ({{\bf Q}_0}) \right ]^2  \left ( [a_{u}(\vec{r};{{\bf Q}_0})]^2 - [b_{1g}(\vec{r};{{\bf Q}_0})]^2  \right ) \\
+ \left [ C_{b_{2g}}^{b_{3u}} ({{\bf Q}_0}) \right ]^2 \left ( [b_{3u}(\vec{r};{{\bf Q}_0})]^2 - [b_{2g}(\vec{r};{{\bf Q}_0})]^2  \right ) .
\end{split}
\end{equation}
The transition densities are obtained as follows,
\begin{equation}\label{eq:gamma02}
\begin{split}
\gamma_{02}(\vec{r};{{\bf Q}_0}) =  C_{b_{1g}}^{b_{3u}} ({{\bf Q}_0}) \sqrt{2} b_{1g}(\vec{r};{{\bf Q}_0}) b_{3u}(\vec{r};{{\bf Q}_0}) \\
+ C_{b_{2g}}^{a_{u}} ({{\bf Q}_0}) \sqrt{2} b_{2g}(\vec{r};{{\bf Q}_0}) a_{u}(\vec{r};{{\bf Q}_0}) ,
\end{split}
\end{equation}
\begin{equation}\label{eq:gamma03}
\begin{split}
\gamma_{03}(\vec{r};{{\bf Q}_0}) =  C_{b_{1g}}^{a_{u}} ({{\bf Q}_0}) \sqrt{2} b_{1g}(\vec{r};{{\bf Q}_0}) a_{u}(\vec{r};{{\bf Q}_0}) \\
+ C_{b_{2g}}^{b_{3u}} ({{\bf Q}_0}) \sqrt{2} b_{2g}(\vec{r};{{\bf Q}_0}) b_{3u}(\vec{r};{{\bf Q}_0})  ,
\end{split}
\end{equation}
\begin{equation}\label{eq:gamma23}
\begin{split}
\gamma_{23}(\vec{r};{{\bf Q}_0}) =  \left [ C_{b_{2g}}^{b_{3u}} ({{\bf Q}_0}) C_{b_{2g}}^{a_{u}} ({{\bf Q}_0}) +  C_{b_{1g}}^{b_{3u}} ({{\bf Q}_0}) C_{b_{1g}}^{a_{u}} ({{\bf Q}_0})  \right ] \\ 
\times b_{3u}(\vec{r};{{\bf Q}_0}) a_{u}(\vec{r};{{\bf Q}_0}) \\
+ \left [ C_{b_{2g}}^{b_{3u}} ({{\bf Q}_0}) C_{b_{1g}}^{b_{3u}} ({{\bf Q}_0}) +  C_{b_{2g}}^{a_{u}} ({{\bf Q}_0}) C_{b_{1g}}^{a_{u}} ({{\bf Q}_0})  \right ] \\ 
\times b_{2g}(\vec{r};{{\bf Q}_0}) b_{1g}(\vec{r};{{\bf Q}_0}) .
\end{split}
\end{equation}
The molecular (electronic and nuclear) wavepacket reads 
\begin{equation}\label{eq:totalpsi}
\begin{split}
\Psi({\vec{r}}_1,\sigma_1, ... ,{\vec{r}}_{42},\sigma_{42},{\bf Q},t) = \\
 \sum_{s=0,2,3} \psi^{(s)}({\bf Q},t)  {^{1}\mathrm{\Phi_s}}({\vec{r}}_1,\sigma_1, ... ,{\vec{r}}_{42},\sigma_{42};{\bf Q}) .
\end{split}
\end{equation}
As a consequence, the corresponding time-dependent one-electron reduced density at the ground-state equilibrium geometry (FC point), ${\bf Q}={\bf Q}_0$, 
expands as
\begin{equation}\label{eq:reddens}
\begin{split}
\rho_{\Psi}({\vec{r}},{{\bf Q}_0},t) =  \sum_{s=0,2,3} |\psi^{(s)}({{\bf Q}_0},t)|^2 \rho_{ss}({\vec{r}};{{\bf Q}_0}) \\
+ \sum_{s=0,2,3} \sum_{3\geq p > s} 2\Re \{ \psi^{(s)*}({{\bf Q}_0},t) \psi^{(p)}({{\bf Q}_0},t) \} \gamma_{sp} ({\vec{r}};{{\bf Q}_0}) .
\end{split}
\end{equation}
We then introduce the time-dependent differential density, $\delta_{\Psi}({\vec{r}},{{\bf Q}_0},t)$, as follows
\begin{equation}\label{eq:reddens2}
\begin{split}
\rho_{\Psi}({\vec{r}},{{\bf Q}_0},t) = \Big [ \sum_{s=0,2,3} |\psi^{(s)}({{\bf Q}_0},t)|^2 \Big ] \rho_{00}({\vec{r}};{{\bf Q}_0}) + \delta_{\Psi}({\vec{r}},{{\bf Q}_0},t) ,
\end{split}
\end{equation}
where
\begin{equation}\label{eq:diffdens}
\begin{split}
\delta_{\Psi}({\vec{r}},{{\bf Q}_0},t) =  \sum_{s=2,3} |\psi^{(s)}({{\bf Q}_0},t)|^2 \delta_{0s}({\vec{r}};{{\bf Q}_0}) \\
+ \sum_{s=0,2,3} \sum_{3\geq p > s} 2\Re \{ \psi^{(s)*}({{\bf Q}_0},t) \psi^{(p)}({{\bf Q}_0},t) \} \gamma_{sp} ({\vec{r}};{{\bf Q}_0}) .
\end{split}
\end{equation}
It represents the time evolution of the electron-hole pair (positive values: electron excess, negative values: electron depletion). The contributions due to population differences, $\delta_{02}$ and $\delta_{03}$, are $A_g$ (totally symmetric); that due to $\gamma_{02}$ is $B_{2u}$ ($y$-polarised), that due to $\gamma_{03}$ is $B_{1u}$ ($z$-polarised), and that due to $\gamma_{23}$ is $B_{3g}$ ($yz$-component of a quadrupole). Static contributions are all displayed in Fig.\ \ref{fig:transdens}, while the resulting wavepacket is shown in Fig.\ \ref{fig:elecwp}.

For the electronic wavepacket, we examine its shape over the time domain between 11.1 and 12.3 fs (time window of 1.2 fs). The pump pulse is then almost off (duration $\sim$1 fs centred at 10 fs; the effect of its presence is essentially visible between 9 and 11 fs, when the local populations at FC keep varying) but the nuclear motion has not started yet so that the relative local populations are still about constant: see Fig.\ \ref{fig:elecwp}. The relative local coherences oscillate during that time window: about 1.5 periods for the $S_0-S_2$ coherence, about two periods for $S_0-S_3$, about half a period for $S_2-S_3$. As a result, shown in Fig.\ \ref{fig:elecwp}, $\delta_{\Psi}({\vec{r}},{{\bf Q}_0},t)$ exhibits some complicated oscillatory behaviour. 

The electronic wavepacket as such is not an observable. However, its time evolution produces a net displacement of charges over time that create oscillations in the time-dependent dipole moment: see Fig.\ \ref{fig:elecwp}. Due to symmetry, the matrix representation of the dipole moment has only two nonzero components: $\mu_{02}^y({\bf Q}_0)$ and $\mu_{03}^z({\bf Q}_0)$. The corresponding electronic coherences thus induce a time-dependent permanent dipole, which approximately reads 
\begin{equation}\label{eq:tdms}
\begin{split}
\vec{\mu}({{\bf Q}_0},t) \approx  \frac{2 \Re \{ \psi^{(0)*}({{\bf Q}_0},t) \psi^{(2)}({{\bf Q}_0},t) \}  }{\sum_{s=0,2,3}|\psi^{(s)}({{\bf Q}_0},t)|^2} \mu_{02}^y({{\bf Q}_0}) {\vec{e}}_y \\
+ \frac{2 \Re \{ \psi^{(0)*}({{\bf Q}_0},t) \psi^{(3)}({{\bf Q}_0},t) \}  }{\sum_{s=0,2,3}|\psi^{(s)}({{\bf Q}_0},t)|^2} \mu_{03}^z({{\bf Q}_0}) {\vec{e}}_z ,
\end{split}
\end{equation}
where ${\vec{e}}_y$ and ${\vec{e}}_z$ are unit vectors. The exact expression should involve integration over the nuclear coordinates. Such oscillations are expected to be monitored experimentally, thus revealing direct coherences between both $S_0$ anf $S_2$ and $S_0$ and $S_3$. Indirect coherence between $S_2$ and $S_3$ can be inferred from the latter but may also be monitored upon probing the time-dependent quadrupole, since it multiplies $\gamma_{23}$, which is the $yz$-component of a symmetric tensor of rank two.

As was shown here, the simultaneous excitation of $S_2$ and $S_3$ by a single pump pulse is able to produce an electronic wavepacket at early times, once electronic populations are stabilised after the pumping stage, yet before the nuclear motion kicks off.
Below, we shall study the extent to which the nuclear motion will affect the survival of the electronic coherence at longer times.

\vspace{0.5cm}
\leftline{{III.1 COHERENCE SURVIVAL }}
\vspace{0.5cm}

Here, we provide deeper insight into the effect of the nuclear motion on the electronic coherence and its lifetime in a rigid molecule.
Various models will be compared. They all consider a reduced set of nuclear coordinates, the five totally symmetric modes of pyrazine.
They are the degrees of freedom that are excited first, before intramolecular vibrational energy redistribution (IVR) starts taking place significantly towards other modes.
Our simulations are thus expected to be realistic over a typical time scale of about 100 fs.
However, we pushed them up to 1 ps -- as a model study -- in order to assess the robustness over time of two usual approximations: the harmonic one and the Franck-Condon one.
Note that the effect of non totally symmetric modes, with IVR, and how some of them should ultimately induce nonadiabatic transitions will be presented in future work.

Throughout the present work, the molecule was prepared initially in the lowest vibrational eigenstate of the $S_0$ ground electronic state.
More specifically, in the harmonic approximation the initial state was taken as the product of the $|i;\nu=0\rangle$ eigenstates
of the $\Big[ -\frac{1}{2}\frac{\partial^2}{\partial Q_i^2} + \frac{1}{2} k_0^{ii} Q_i^2 \Big]$ operators ($i=1, 2, 3, 4, 5$) where the $k_0^{ii}$'s are the diagonal elements of the ${\bf k}_0$ matrix (see SI). On the other hand, in the anharmonic case the initial state was built up as the product of the $|i;\nu=0\rangle$ eigenstates
of the
$\Big[ -\frac{1}{2}\frac{\partial^2}{\partial Q_i^2} + D_0^i \bigg( 1 - e^{\epsilon_i \sqrt{\frac{k_0^{ii}}{2 D_0^i }} Q_i}  \bigg)^2 \Big]$
operators ($i=1, 2, 3, 4, 5$) where $k_0^{ii}$'s are defined above and $D_0^i$ is the $i^{th}$ element of the ${\bf D}_0$ vector while $\epsilon_i = \pm 1$ according to the oriention of each mode with respect to its dissociation asymptote (see SI).

\begin{figure}[ht]
\includegraphics[clip,width=8.8cm]{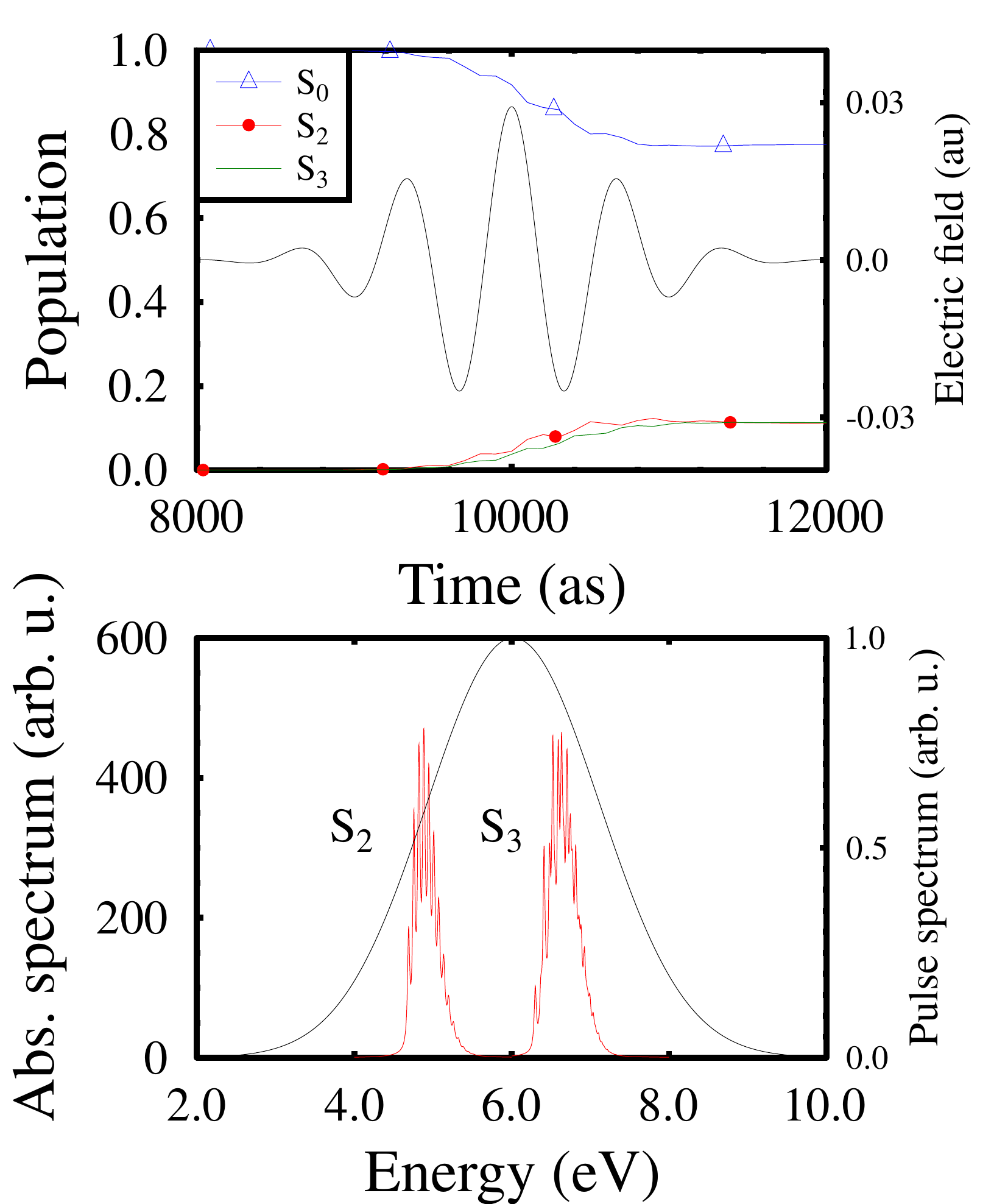} \caption{
(upper panel) Impulsive excitation of the $S_2$ and $S_3$ electronic states of the pyrazine molecule in the H+FC model. For the parameter values of the applied pump pulse (eq.\ \ref{eq:pump}), see the text. (lower panel) Absorption spectrum of the $S_2$ and $S_3$ states calculated through eq.\ \ref{eq:Sp} ($\tau = 40$ fs) in the harmonic model. Here the spectrum of the pump pulse depicted in the upper panel is shown by the solid black line.
}
\label{fig:pumping} 
\end{figure}

In order to achieve significant population transfer to both $S_2$ and $S_3$ from the ground state, an impulsive excitation was applied by setting $t_p = 1$ fs and 
$I_0 = 3\times 10^{13}$ W/cm$^2$.
The central frequency of the pump was tuned in between the $S_2$ and $S_3$ bands ($\hbar\omega_p = 6.02$ eV) as shown by the vertical dashed 
line in Fig.\ \ref{fig:spectra2} (note that the laser polarization direction is $\theta = 45^{\circ}$ (Fig.\ \ref{fig:system})).
For such a short pulse duration the pump pulse was shown to have a wide enough bandwidth to cover the $S_2$ and $S_3$ bands resulting in a simultaneous excitation of both excited states.
This is illustrated in Fig.\ \ref{fig:pumping} on the H+FC model, but we did not notice any significant difference among the four models on that time scale.
The electronic populations after 12 fs ($\sim$ 0.1) do no longer change up to the end of our simulations (1 ps).
However, coherences keep oscillating for the whole duration, as will be shown below.

\begin{figure}[ht]
\includegraphics[clip,width=9.8cm]{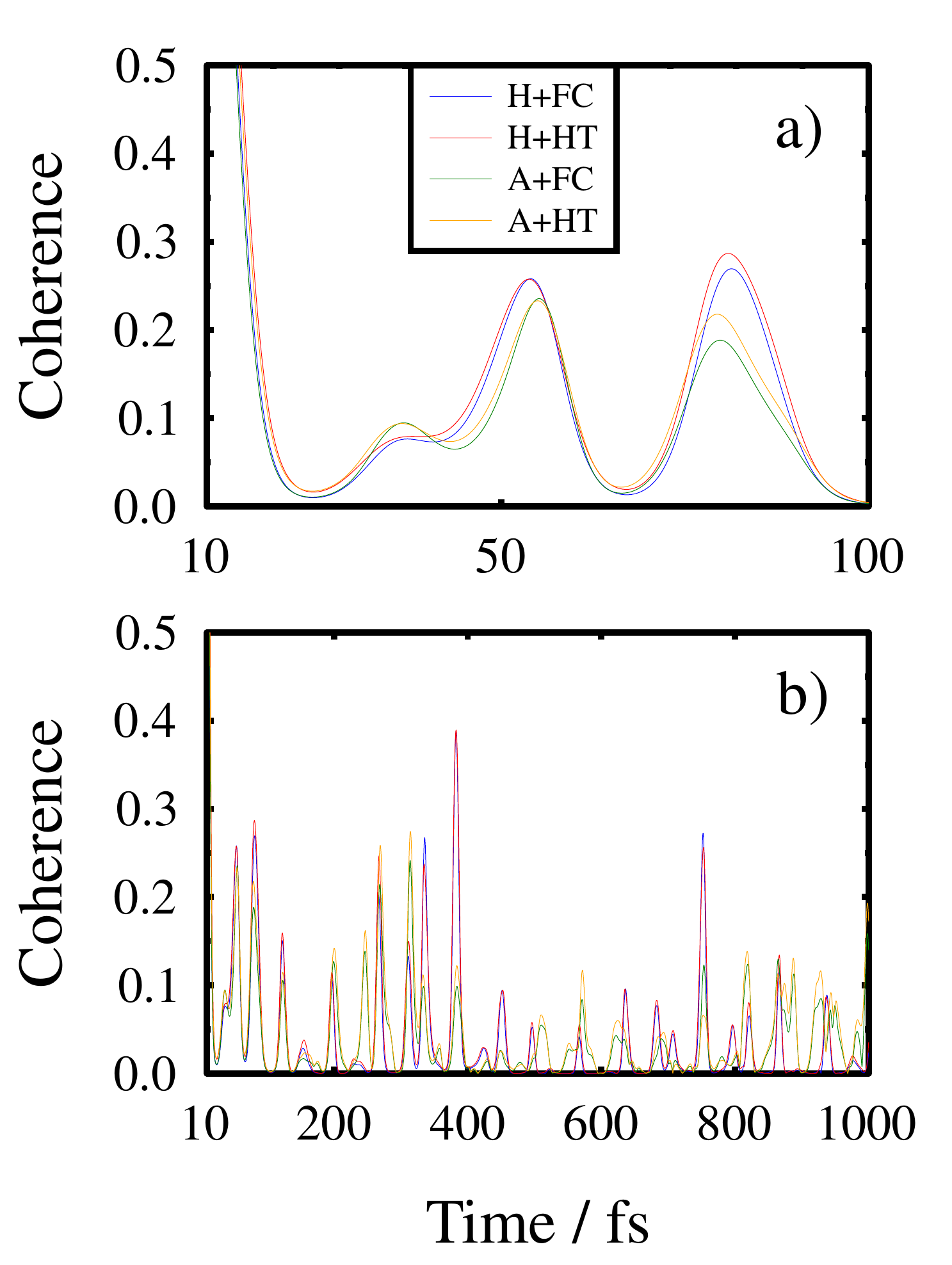} \caption{
Time evolution of the modulus of the $S_2 - S_3$ electronic coherence calculated via eq.\ \ref{eq:Co} for the four models of the pyrazine molecule.}
\label{fig:coh23} 
\end{figure}

As mentioned above, electronic populations remain constant over time once the pump pulse is off. However, the coherences oscillate, which is expected to produce an electronic wavepacket.
This is shown in Fig.\ \ref{fig:coh23} on the four models for the modulus of the function $C(t)$, which is a measure of the coherence between $S_2$ and $S_3$ according to eq.\ \ref{eq:Co}.
On the upper panel (up to 100 fs), we can observe similar temporal behaviours for the four models. There is a pseudoperiod of about 20 fs, which is consistent with the typical time scale of vibrations.
Interestingly enough, the amplitude of the oscillations is similar for either H+FC and H+HT or A+FC and A+HT.
In other words, anharmonicity has significant impact, while the coordinate-dependence of the TDMs is of little influence.
On this time scale, for which a reduced model is realistic, we can conclude that H+FC is overall a good approximation.
For improving it, one should start with anharmonicity.
From a spectral point of view, conclusions are somewhat reversed.
The harmonic approximation has an effect on the energies of the highest vibrational peaks but not so much on their intensities.
In contrast, intensities depend significantly on the coordinate-dependence of the TDMs for the $S_3$ state (see Fig.\ \ref{fig:spectra2}).

Now, let us examine what occurs at longer times. The simulations are no longer realistic for the pyrazine molecule \emph{per se}.
They are to be considered as a model study that will give insight into the possible destruction of electronic coherence when nuclei move in a rigid molecule. 
The largest value (not shown) in Fig.\ \ref{fig:coh23} is almost 1 at 10 fs when pumping is maximal. Early recurrences (before 100 fs) reach values of about 0.3.
As can be observed on the lower panel, long-term recurrences can still reach such significant values at very long times, more so for the harmonic models (even 0.4 at 400 fs, 0.3 again at 750 fs).
The more realistic anharmonic models experience larger damping but still provide values of about 0.15 at such times.
They are subject to stronger dephasing among vibrational oscillations by construction. However, this has little effect on the survival of electronic coherence.
The latter keeps a similar temporal structure, only slightly more damped. 

\begin{figure}[ht]
\includegraphics[clip,width=8.8cm]{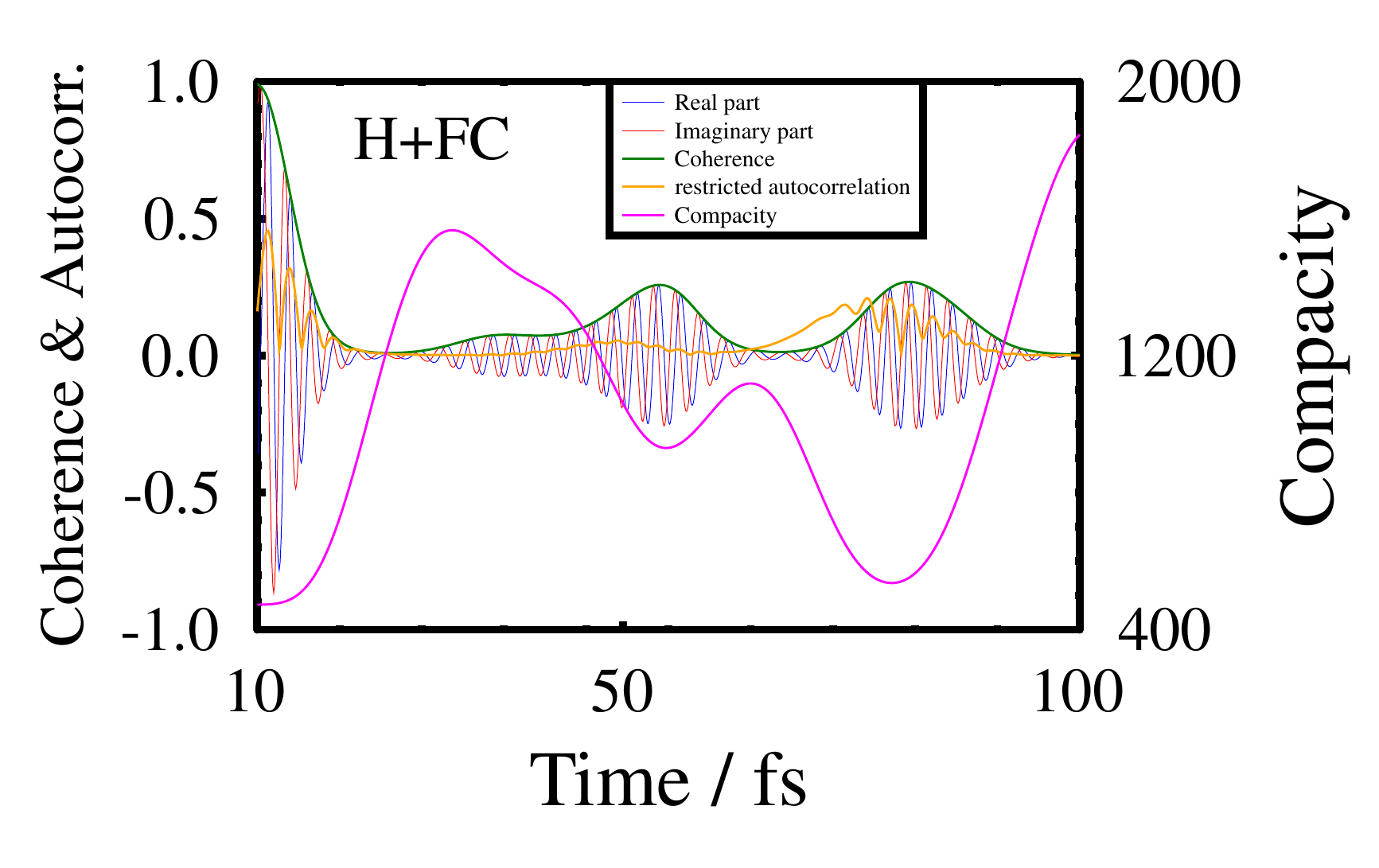} \caption{
Time evolution of the coherence between $S_2$ and $S_3$ compared to the restricted autocorrelation function and to the compacity in the H+FC model.
}
\label{fig:autocohcomp} 
\end{figure}

Finally, let us now determine what in the time evolution of the total wavepacket may affect the electronic coherence.
We first examined whether the autocorrelation function $A(t)$ (eq.\ \ref{eq:Ac}) was correlated to the modulus of $C(t)$ (eq.\ \ref{eq:Co}).
Nothing could be extracted from that comparison because the time dependence is then dominated by what was left in $S_0$ ($\sim$ 80$\%$ of the total wavepacket).
In order to increase contrast and better characterise excited-state dynamics, we defined a restricted autocorrelation function, $A^{\mathrm{(R)}}(t)$, according to (eq.\ \ref{eq:Ar}),
as if nothing remained in $S_0$ after pumping (impulsive picture with half on $S_2$ and half on $S_3$).
This quantity exhibits similar oscillation patterns compared to the modulus of $C(t)$, as shown for the H+FC model in Fig.\ \ref{fig:autocohcomp}.
This seems to indicate that electronic coherence can reconstruct itself when the wavepacket comes back to its original position, as expected from a qualitative perspective.
Similar results were obtained for the other three models and on longer times.

\begin{figure}[ht]
\includegraphics[clip,width=8.8cm]{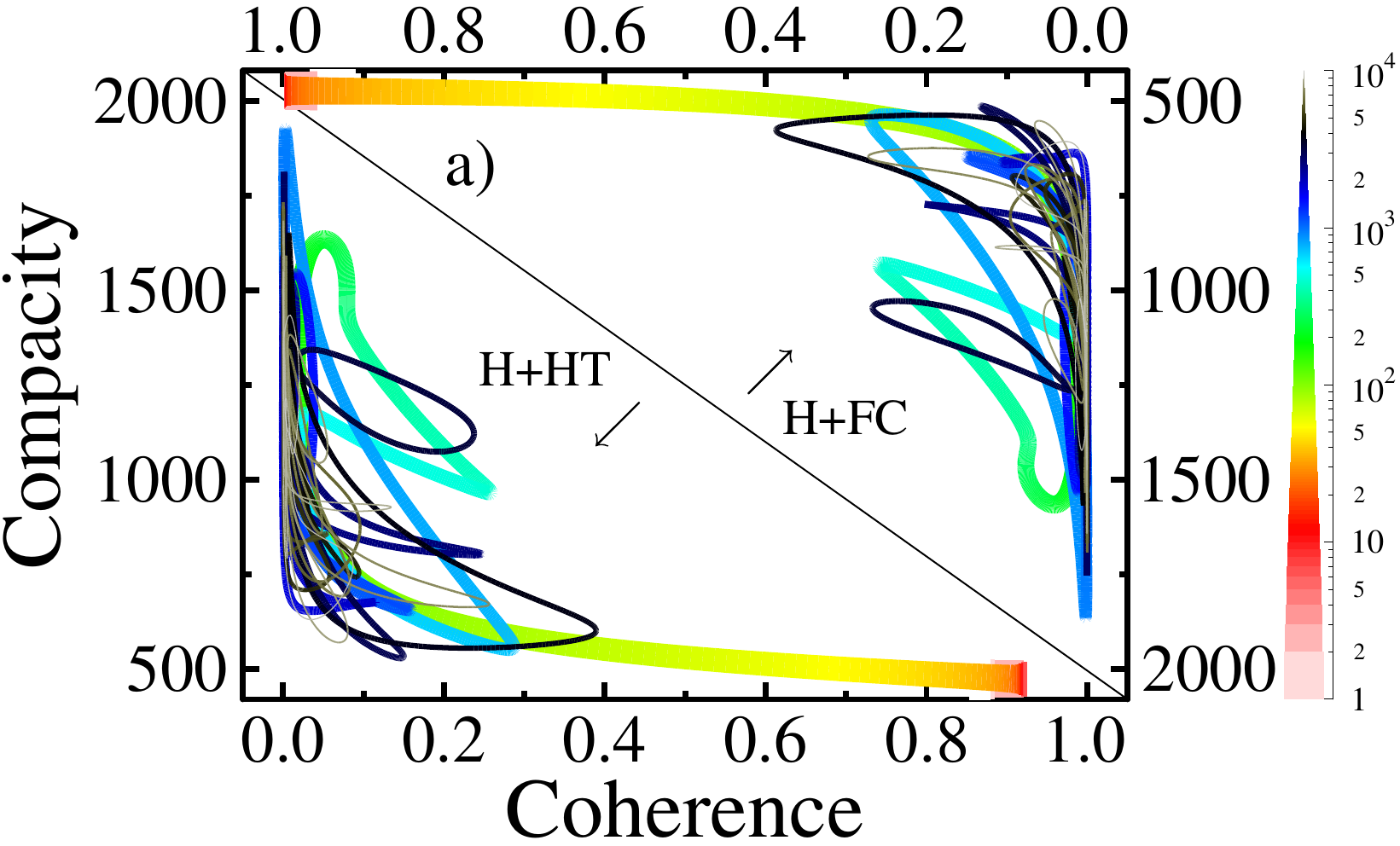} \\
\includegraphics[clip,width=8.8cm]{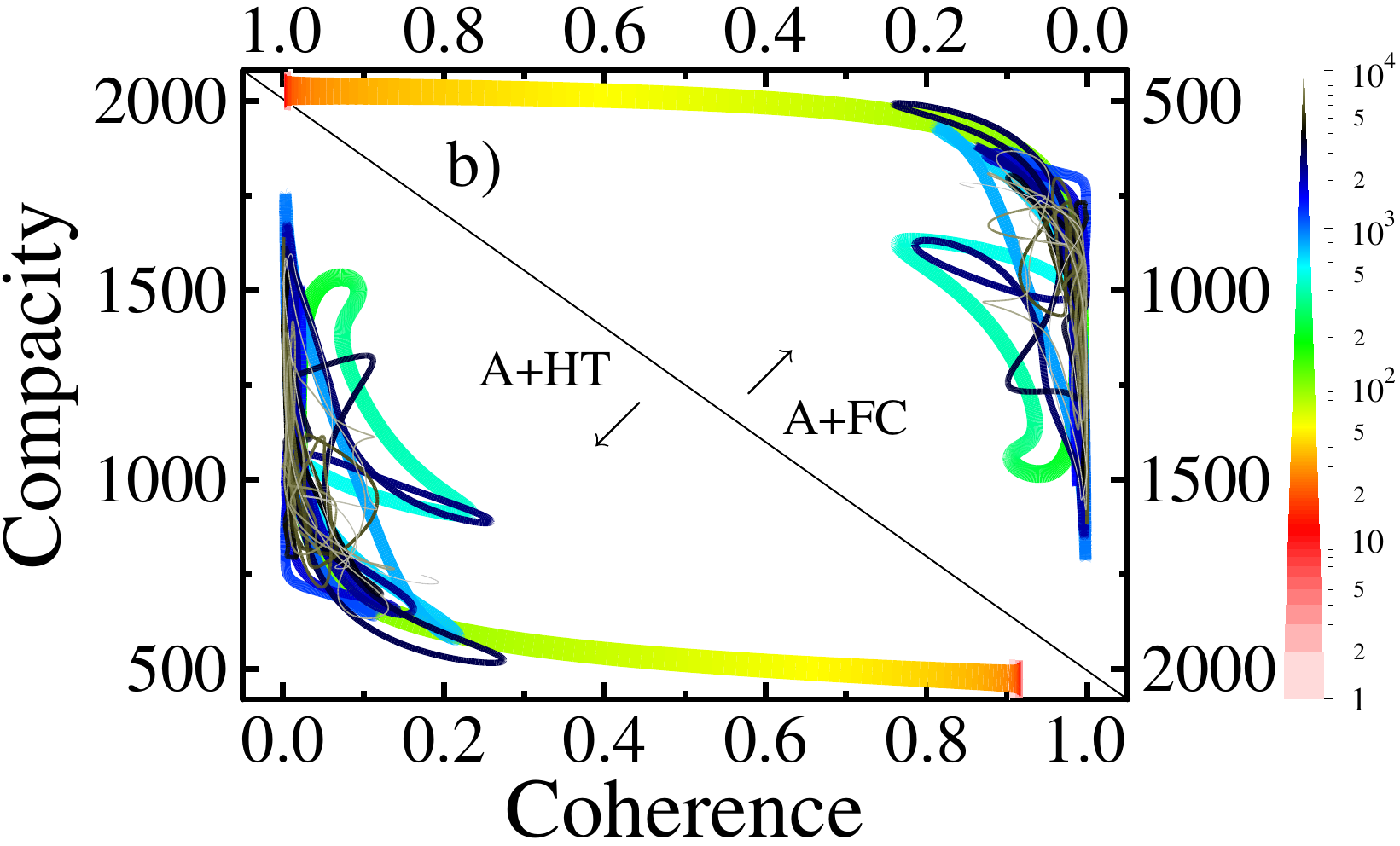} \caption{
Wavepacket compacity against modulus of the coherence between $S_2$ and $S_3$ in the four models (reversed axes were used for better comparison).
Colour and thickness is related to time evolution.
}
\label{fig:cohcomp} 
\end{figure}

The previous observation is consistent with the fact that the electronic coherence is dominated by overlapping properties.
This implies that wavepackets must return to similar regions in the space of nuclear coordinates but also that their dispersion should stay moderate.
A quantitative measure of spreading in space is provided by the compacity function $K(t)$ (eq.\ \ref{eq:Cp}).
As can be observed in Fig.\ \ref{fig:autocohcomp}, $K(t)$ and $C(t)$ exhibit similar temporal oscillations, but their correlation could be accidental over this time window.
A more systematic comparison is provided on long time scales for the four models (see Fig.\ \ref{fig:cohcomp}).
Typically, they show that a large coherence -- especially at early times -- goes with a small compacity, whereas a large compacity rather goes with a small coherence.
Harmonic models (upper panel) are less conclusive on this front, since they allow coherences to take larger values on longer times.
However, such a correlation is more striking in the anharmonic models (lower panel) that experience more phase dispersion of the wavepacket by construction.

\vspace{0.5cm}
\leftline {\bf IV. CONCLUSIONS}
\vspace{0.5cm}
We presented various models of pyrazine to investigate the lifetime of electronic coherence in a rigid and neutral molecule after simultaneous excitation of a pair of bright electronic states by the same pump pulse.
The effect of the Franck-Condon approximation (constant TDMs) on the intensities of the absorption spectra is significant but shows little impact on the survival of electronic coherence.

On the other hand, anharmonicity compared to a harmonic model shows some effect on the electronic coherence but will not destroy it over a time scale of about 1 ps. 
From our results, the survival of electronic coherences seems dominated by the capability of the nuclear wavepacket to exhibit temporal recurrences so as to reconstruct itself regularly both in terms of location and dispersion in space.

Perhaps the strongest approximation in this context, which was not lifted here, is the adiabatic (Born-Oppenheimer) one.
In future work, we shall examine with a full-dimensional model (24 vibrational modes) the effect of breaking the original molecular symmetry along nontotally symmetric modes, which may, of course, induce IVR, but more crucially, nonadiabatic couplings and nonradiative decay among electronic states, including dark ones, via conical intersections. 

\vspace{0.5cm}
\leftline {\bf ACKNOWLEDGEMENTS}
\vspace{0.2cm}
This research was supported by the EU-funded Hungarian grant EFOP-3.6.2-16-2017-00005. 
The authors are grateful to NKFIH for support (Grant K128396).


\end{document}